\documentclass[10pt,a4paper]{article}
\date{}
\usepackage{float}
\usepackage{graphicx}
\usepackage[english]{babel}
\usepackage{amssymb}
\usepackage{amsmath}

\title{Vacuum Energy of a Non-relativistic String with Nontrivial Boundary Condition}
\author{A. Jahan$^1$ and S. Sukhasena$^2$\\
$^1$\emph{Research Institute for Astronomy and Astrophysics of Maragha, Iran}
\\\emph{jahan@riaam.ac.ir}\\
$^2$\emph{The Institute for Fundamental Study,"The Tah Poe Academia", Naresuan University}\\
\emph{Phitsanulok 65000, Thailand}\\
\emph{secksons@nu.ac.th}}
\begin{document}
\maketitle
\begin{abstract}
We derive the partition function of a non-relativistic quantum string which its ends are allowed to freely slide on the two angled straight solid rods. We first derive the classical solution of the model and then use it to derive the partition function utilizing the path integral method. We show that the vacuum energy is sum of the L\"{u}scher potential and a term which depends on the relative angle between rods.
\end{abstract}
\section{Introduction}
In superstring theory, a $Dp$-brane is a $p$-dimensional subspace of bulk space on which an open superstring can end [1]. An open string can end on a same $D$-brane or on different $D$-branes of different dimensions. The later setup is called a $Dp$-$Dp'$ system. In general, there may be relative angles between the $D$-branes. Exchange of closed strings between the $D$-branes gives rise to an attractive force between them. But, this is equivalent to one-loop vacuum amplitude (Casimir energy or vacuum energy) of an open string ending on $D$-branes. Therefore, in supertring theory the Casimir energy of a string manifests itself as the attractive force between the different $Dp$-branes [1].\\
The Casimir energy of a relativistic string has been also studied in quantum chromodynamics. For a static quark-antiquark pair the chromoelectic field can be effectively described by a vibrating string. Then the potential between the pair reveals itself as the ground-state energy of the string, known as L\"{u}scher potential [9-12].\\
In addition to the above-mentioned studies, the vacuum energy of the open and closed strings has been investigated by several authors form different perspectives. The Casimir energy of a closed non-relativistic string and its generalization to a piecewise string is considered by Brevik et al [12, 14-16]. The vacuum energy of an open string placed between two beads is calculated by D'hoker et al [17]. The L\"{u}scher potential is recovered when the masses of beads become large. The quantum corrections to the L\"{u}scher potential was obtained by Elizalde et al [18]. They interpreted the corrections as a sort of non-local effect in a Bosonic string. The Nambu-Goto model of an open string was used by Kleinert et al to model the inter-quark potential. In their work, the string is assumed to end on point masses with mass $m$. They showed that one recovers the L\"{u}scher potential in the limits $m\rightarrow 0$ or $m\rightarrow \infty$ [19, 20]. Possible implications of the vacuum energy of a closed string on the cosmological scales have been investigated by Kikkawa et al [21] and Sendai et al [22].  \\
Motivated by the aforementioned studies, in the present work we consider a non-relativistic string which its ends are located on two straight solid rods (See Fig.1). The ends of string are allowed to freely slide on the rods. There is an interesting analogy between this problem and $D1$-$D1$ system in superstring theory [2-4]. Here, the rods play the role of $D1$-branes and like the $D1$-$D1$ system, the classical solutions depend on the relative angle between the rods. The main difference is that here we are dealing with a non-relativistic string. The boundary conditions depends on the relative angle between the rods. For example, the parallel and anti-parallel rods give rise to the Neuman-Neuman and Dirichlet-Dirichlet boundary conditions. The Neuman-Dirichlet (mixed) boundary condition arises when the angle is $\frac{\pi}{2}$. Using a method based on the path integral formalism [3-7], we derive the partition function of the quantized string. Yet, like the $D1$-$D1$ system the result depends on the angle between the rods. Finally, we obtain the Casimir energy as the zero-temperature limit of the free energy.
\section{Classical Dynamics}
Let us consider a non-relativistic string which its ends freely move along two straight solid rods. One of the rods is located at $z=0$ and the other at $z=l$. The angles between the $X$-axis and the rods at $z=0$ and and $z=l$ are denoted by $\theta_1$ and  $\theta_2$, respectively. The displacement of string from equilibrium lies on the $X-Y$ plane and can be described by the displacement field  $\vec\phi(z,t)$, which can be written as
\begin{equation}
\vec\phi(z,t)=\phi_x(z,t)\widehat{e}_x+\phi_y(z,t)\widehat{e}_y.
\end{equation}
where $\widehat{e}_x$ and $\widehat{e}_y$ are the unit vectors along the $X$-axis and $Y$-axis, respectively. The potential energy of an infinitesimal segment of a taut string with length $l$ is the work done to stretched the segment under the constant tension $T$. Thus, for the whole string one writes [7]
{\setlength\arraycolsep{2pt}
\begin{eqnarray}
\nonumber
V&=&T\int_0^ldz\bigg[\sqrt{1+\bigg(\frac{\partial\vec\phi(z,t)}{\partial z}\bigg)^2}-1\bigg]\\
&\approx&\frac{1}{2}T\int_0^ldz\bigg(\frac{\partial\vec\phi(z,t)}{\partial z}\bigg)^2.
\end{eqnarray}}
provided that $|\vec\phi|\ll 1$. The kinetic energy is [7]
\begin{equation}\label{2}
K=\frac{1}{2}\mu A\int_0^ldz\bigg(\frac{\partial\vec\phi(z,t)}{\partial t}\bigg)^2.
\end{equation}
where $A$ is the string cross section and $\mu$ denotes its mass density. Therefore, the action becomes
\begin{equation}\label{3}
S=\frac{1}{2}\int^{t_2}_{t_1}dt\int_0^l\mu A\bigg(\frac{\partial\vec\phi(z,t)}{\partial t}\bigg)^2-T\bigg(\frac{\partial\vec\phi(z,t)}{\partial z}\bigg)^2dz.
\end{equation}
The equation of motion is given by
\begin{equation}\label{1}
\frac{\delta S}{\delta\phi_i}=0.
\end{equation}
where the variation of action can be obtained by means of an integration by part. Hence, we have
\begin{figure}[t]
\centering
  \includegraphics[width=70mm]{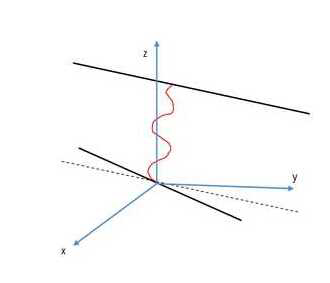}
  \caption{A taut string which its ends can move freely along the solid rods. The rods are at relative angle.}
\end{figure}
{\setlength\arraycolsep{2pt}
\begin{eqnarray}
\nonumber
\delta S&=&\int^{t_2}_{t_1}dt\int_0^l\sum_{i=1}^2\delta\phi_i\bigg(\mu A\frac{\partial^2\phi_i(z,t)}{\partial t^2}-T\frac{\partial^2\phi_i(z,t)}{\partial z^2}\bigg)dz\\
&+&\mu A\sum_{i=1}^2\delta\phi_i(z,t)\frac{\partial\phi_i(z,t)}{\partial t}\bigg{|}_{t_1}^{t_2}-T\sum_{i=1}^2\delta\phi_i(z,t)\frac{\partial\phi_i(z,t)}{\partial z}\bigg{|}_{0}^{l}.
\end{eqnarray}}
So from (5) and (6), provided that the boundary terms vanish, the equation of motion becomes
\begin{equation}\label{1}
\frac{1}{v^2}\frac{\partial^2\vec\phi(z,t)}{\partial t^2}-\frac{\partial^2\vec\phi(z,t)}{\partial z^2}=0.
\end{equation}
where the speed of sound is $v=\big(\frac{T}{\mu A}\big)^{\frac{1}{2}}$. The ends of string satisfy the following set of constraints
{\setlength\arraycolsep{2pt}
\begin{eqnarray}
\phi_2(0,t)-\tan\theta_1 \phi_1(0,t)&=&0,\\
\phi_2(l,t)-\tan\theta_2 \phi_1(l,t)&=&0.
\end{eqnarray}}
Taking into account the above constraints, the third term of (6) vanishes, provided that
{\setlength\arraycolsep{2pt}
\begin{eqnarray}
\frac{\partial\phi_1(0,t)}{\partial z}+\tan\theta_1\frac{\partial\phi_2(0,t)}{\partial z}&=&0,\\
\frac{\partial\phi_1(l,t)}{\partial z}+\tan\theta_2\frac{\partial\phi_2(l,t)}{\partial z}&=&0.
\end{eqnarray}}
To solve the wave equation, we set  $\phi_i(z,t)=f(t)g_i(z)$ in (7), which leads to separation of the wave equation as
{\setlength\arraycolsep{2pt}
\begin{eqnarray}
\frac{d^2f(t)}{dt^2}+\omega^2f(t)&=&0,\\
\frac{d^2g_i(z)}{dz^2}+k^2g_i(z)&=&0.
\end{eqnarray}}
where the dispersion relation admits $\omega=vk$. The second equation has the solution of the form
\begin{equation}\label{1}
g_i(z)=a_ie^{i\beta}e^{ikz}+\bar{a}_ie^{-i\beta}e^{-ikz}.
\end{equation}
Substituting (14) in boundary conditions (10) and (11), gives rise to
{\setlength\arraycolsep{2pt}
\begin{eqnarray}
\tan\theta_1&=&-\frac{a_1e^{i\beta}-\bar{a}_1e^{-i\beta}}{a_2e^{i\beta}-\bar{a}_2e^{-i\beta}},\\
\tan\theta_2&=&-\frac{a_1e^{i\beta}e^{ikl}-\bar{a}_1e^{-i\beta}e^{-ikl}}{a_2e^{i\beta}e^{ikl}-\bar{a}_2e^{-i\beta}e^{-ikl}}.
\end{eqnarray}
Choosing the constants as
{\setlength\arraycolsep{2pt}
\begin{eqnarray}
a_1&=&\bar{a}_1=a,\\
a_2&=&-\bar{a}_2=-ia,
\end{eqnarray}}
modify (15) and (16) to
{\setlength\arraycolsep{2pt}
\begin{eqnarray}
\tan\theta_1&=&\tan\beta,\\
\tan\theta_2&=&\tan(\beta+k l),
\end{eqnarray}}
and (14) to
{\setlength\arraycolsep{2pt}
\begin{eqnarray}
g_1(z)&=&a\cos(kz+\beta),\\
g_2(z)&=&a\sin(kz+\beta).
\end{eqnarray}}
Equations (19) and (20) can be solved to yield
{\setlength\arraycolsep{2pt}
\begin{eqnarray}
\theta_1&=&\beta+n_1\pi,\\
\theta_2&=&\beta+kl+n_2\pi.
\end{eqnarray}}
with $n_1,n_2\in \mathbb{Z}$. These relations combine to yield the quantized wave number $k_n=\frac{n_r\pi }{l}$. Therefore, from (21)-(24) we find
{\setlength\arraycolsep{2pt}
\begin{eqnarray}
g_1(z)&=&a_n\cos\Big(\frac{n_r\pi z}{l}+\theta_1+n_1\pi\Big),\\
g_2(z)&=&a_n\sin\Big(\frac{n_r\pi z}{l}+\theta_1+n_1\pi\Big).
\end{eqnarray}}
where we have defined $\theta_2-\theta_1=\pi r$ and $n_r=n+r$ with $0\leq r\leq1$. So, from (25), (26) and (12) we obtain the general solutions as
{\setlength\arraycolsep{2pt}
\begin{eqnarray}
\phi_1(z,t)&=&\sum_{n=-\infty}^\infty \Big(a_ne^{i\omega_nt}+\overline{a}_ne^{-i\omega_nt}\Big)\cos\Big(\frac{n_r\pi z}{l}+\theta_1\Big),\quad\omega_n=vk_n \\
\phi_2(z,t)&=&\sum_{n=-\infty}^\infty \Big(a_ne^{i\omega_nt}+\overline{a}_ne^{-i\omega_nt}\Big)\sin\Big(\frac{n_r\pi z}{l}+\theta_1\Big).
\end{eqnarray}}
The factor $n_1\pi$ in (25) and (26) is now absorbed in expansion coefficients. From now on we set $\theta_1=0$ which is achieved by an appropriate rotation of system of coordinates.
\section{Partition Function}
One way to drive the vacuum energy (ground-state energy) is to take the zero-temperature limit of free energy, namely
\begin{equation}\label{1}
E_0=\lim_{\beta\rightarrow \infty}F(\beta),
\end{equation}
where the free energy is
\begin{equation}\label{1}
F(\beta)=\lim_{\beta\rightarrow \infty}-\frac{1}{\beta}\ln Z(\beta).
\end{equation}
The partition function at temperature $T={\beta^{-1}}$ can be defined in terms of the path integral as
\begin{equation}\label{8}
 Z(\beta)=\int D\phi_1D\phi_2 e^{-S_E[\vec\phi\,]}.
\end{equation}
The Euclidean action is obtained by Wick rotation of the time parameter upon setting $\tau=it$. The fluctuating fields obey the periodic boundary condition
\begin{equation}\label{1}
\phi_i(\tau+\beta,z)=\phi_i(\tau,z).
\end{equation}
So, the Euclidean action can be written
\begin{equation}\label{7}
S_E=\frac{1}{2}\sum_{i=1}^2\int_0^\beta d\tau\int_0^l dz\,\phi_i\Box_E \phi_i+\textrm{vanishing\,\,boundary\,\,terms},
\end{equation}
where
\begin{equation}\label{1}
\Box_E=\frac{\partial^2}{\partial \tau^2}+\frac{\partial^2}{\partial z^2}.
\end{equation}
The fluctuating fields can be expanded in terms of the eigen-modes $u_{nm}$ and $v_{nm}$
{\setlength\arraycolsep{2pt}
\begin{eqnarray}
\phi_1&=&\sum_{n,m=-\infty}^\infty\frac{\chi_{nm}}{\sqrt2}u_{nm},\\
\phi_2&=&\sum_{n,m=-\infty}^\infty\frac{\chi_{nm}}{\sqrt2}v_{nm}.
\end{eqnarray}}
Since the field components are real we have $\overline{\chi}_{nm}=\chi_{n,-m}$. The eigen-modes assumed to be
{\setlength\arraycolsep{2pt}
\begin{eqnarray}
u_{nm}&=&\cos\Big( \frac{n_r\pi z}{l}\Big) e^{i\omega_m\tau},\qquad \omega_m=\frac{2\pi m}{\beta},\\
v_{nm}&=&\sin\Big(\frac{n_r\pi z}{l}\Big)e^{i\omega_m\tau}.
\end{eqnarray}}
These eigen-modes nullify the boundary terms in (33) and satisfy the eigen-value equations
{\setlength\arraycolsep{2pt}
\begin{eqnarray}
\Box_E u_{nm} =\lambda^r _{nm}u_{nm},\\
\Box_E v_{nm} =\lambda^r _{nm}v_{nm},
\end{eqnarray}}
where the common eigen-value is
\begin{equation}\label{1}
\lambda^r_{nm}=\nu_r^2+\Big(\frac{\omega_m}{v}\Big)^2,\qquad \nu_r=\frac{n_r\pi}{l}.
\end{equation}
One easily observes $\lambda^r_{-n,m}=\lambda^{-r}_{nm}$. After a little algebra one obtains
{\setlength\arraycolsep{2pt}
\begin{eqnarray}\label{27-30}
\int^{\beta}_{0}d\tau\int _0^ldz\, u_{n^\prime m^\prime}u_{nm}&=&\frac{1}{2}\beta l \Bigg(\delta_{n n^\prime}+\frac{(-1)^{n+n^\prime}}{\pi}
\frac{\sin2\pi a}{n+n^\prime+2a}\Bigg)\delta_{m+m^\prime,0},\\
\int^{\beta}_{0}d\tau\int _0^ldz\, v_{n^\prime m^\prime}v_{nm}&=&\frac{1}{2}\beta l \Bigg(\delta_{n n^\prime}-\frac{(-1)^{n+n^\prime}}{\pi}
\frac{\sin2\pi a}{n+n^\prime+2a}\Bigg)\delta_{m+m^\prime,0}.
\end{eqnarray}}
So, with the help of equations (39) - (43) we find for the action
{\setlength\arraycolsep{2pt}
\begin{eqnarray}\label{31}
\nonumber
S_E&=&\frac{1}{2}\sum_{i=1}^2\int_0^\beta d\tau\int_0^l dz\,\phi_i\Box_E \phi_i\\
&=&\frac{1}{4}\beta l\sum_{n,m=-\infty}^{\infty}\chi_{nm}\lambda^r_{nm}\overline{\chi}_{nm}.
\end{eqnarray}}
Therefore, upon invoking
\begin{equation}\label{1}
\int D\phi_1D\phi_2 =\int^{\infty}_{-\infty}\prod_{n,m=-\infty}^{\infty} d{\chi}_{nm},
\end{equation}
the partition function takes the form
{\setlength\arraycolsep{2pt}
\begin{eqnarray}\label{32}
\nonumber
Z(\beta)&=&\int^{\infty}_{-\infty}\prod_{n,m=-\infty}^{\infty} d{\chi}_{nm} \exp\Big(-\frac{1}{4}\beta l\sum_{n,m=-\infty}^{\infty}\chi_{nm}\lambda^r_{nm}\overline{\chi}_{nm}\Big)\\
&=&\prod_{n,m=-\infty}^{\infty}\frac{1}{(\lambda^r_{nm})^{\frac{1}{2}}},
\end{eqnarray}}
where we have utilized the Gaussian integral
\begin{eqnarray}
\int d\xi \,d\overline{\xi}\,e^{-\alpha\xi \overline{\xi}}&=&\frac{2\pi }{\alpha}.
\end{eqnarray}
By noting $\lambda^r_{n,-m}=\lambda^r_{nm}$, the partition function (46) can be recast into
\begin{equation}\label{1}
Z(\beta)={\frac{1}{\sqrt{\lambda^r_{00}}}}\prod_{m=1}^{\infty}\frac{1}{\lambda^{r}_{0m}}
\prod_{n=1}^{\infty}\bigg({\frac{1}{\sqrt{\lambda^{r}_{n0}}}}\prod_{m=1}^{\infty}\frac{1}{\lambda^{r}_{nm}}\bigg)
\prod_{n=1}^{\infty}\bigg({\frac{1}{\sqrt{\lambda^{-r}_{n0}}}}\prod_{m=1}^{\infty}\frac{1}{\lambda^{-r}_{nm}}\bigg).
\end{equation}
The last two term in the above expression can be further simplified to
{\setlength\arraycolsep{2pt}
\begin{eqnarray}\label{32}
\nonumber
\prod_{m=1}^{\infty}\frac{1}{\lambda^{\pm r}_{nm}}&=&\prod_{m=1}^{\infty}\frac{v^2}{\omega^2_{m}}\prod_{m=1}^{\infty}\frac{1}{1+\frac{v^2\nu^2_{\pm r}}{\omega^2_{m}}}\\
&=&\frac{\sqrt{\lambda^{\pm r}_{n0}}}{2\sinh\frac{\beta v\nu_{\pm r}}{2}}.
\end{eqnarray}}
Similarly, we obtain
{\setlength\arraycolsep{2pt}
\begin{eqnarray}\label{32}
\prod_{m=1}^{\infty}\frac{1}{\lambda^{r}_{0m}}
&=&\frac{\sqrt{\lambda^{ r}_{00}}}{2\sinh\frac{\beta vr}{2}},
\end{eqnarray}}
where we have gained the formulas [8]
{\setlength\arraycolsep{2pt}
\begin{eqnarray}
\prod_{m=1}^{\infty}\frac{1}{cm^2}&=&\frac{\sqrt{c}}{2\pi},\\
  \frac{\sinh\pi x}{\pi x}&=& \prod_{m=1}^{\infty}\bigg(1+\frac{x^2}{m^2}\bigg).
\end{eqnarray}}
When (49) and (50) are substituted in (48), we arrive at
{\setlength\arraycolsep{2pt}
\begin{eqnarray}\label{32}
Z(\beta)&=& \frac{1}{2\sinh\frac{\beta\pi v r}{2l}}\bigg(\prod_{n=1}^{\infty}\frac{1}{2\sinh\frac{\beta v\nu_r}{2}}\bigg)\bigg(\prod_{n=1}^{\infty}\frac{1}{2\sinh\frac{\beta v\nu_{-r}}{2}}\bigg).
\end{eqnarray}}
Then, by rearranging the two and third terms of (53) as
\begin{equation}\label{1}
\prod_{n=1}^{\infty}\frac{1}{2\sinh\frac{\beta v\nu_{\pm r}}{2}}=q^{\frac{1}{2}\sum_n(n\pm r)}\prod_{n=1}^{\infty}\frac{1}{1-q^{n\pm r}},\quad\qquad q=e^{-\frac{\pi\beta}{l}v}
\end{equation}
and using the formula (see appendix A)
{\setlength\arraycolsep{2pt}
\begin{eqnarray}\label{33}
\sum^{\infty}_{n=1}(n+r)=\frac{1}{24}-\frac{1}{2}\bigg(r+\frac{1}{2}\bigg)^2,
\end{eqnarray}}
we are left with the final form of the partition function [4]
\begin{equation}\label{1}
Z(\beta)=\frac{q^{-\frac{1}{2}r(r-1)-\frac{1}{12}}}{1-q^r}\prod_{n=1}^{\infty}\frac{1}{1-q^{n+ r}}\frac{1}{1-q^{n- r}}.
\end{equation}
Note that we have discarded an irrelevant divergent term in (55) (see appendix).
\section{Vacuum Energy}
From (29) and (30), we can have the vacuum energy as
{\setlength\arraycolsep{2pt}
\begin{eqnarray}\label{32}
\nonumber
E&=& \lim_{\beta\rightarrow \infty}-\frac{1}{\beta}\ln Z(\beta)\\
&=&-\frac{1}{12}\frac{\pi v}{l}-\frac{1}{2}r(r-1)\frac{\pi v}{l}.
\end{eqnarray}}
The first term of (57) is the L\"{u}scher potential which was calculated for the potential between the heavy quarks within the context of string theoretical considerations in hadrons physics. For a quark located at $z=0$ and an anti-quark located at $z=l$, at long-wavelength limit one can approximate the thin flux tube as a string between the quark and anti-quark. The vacuum energy of such a string manifests itself as the L\"{u}scher potential [9-12].\\
From (57) we observe that the vacuum energy of the string coincides with L\"{u}scher potential when the rods are parallel ($\theta=0$) or anti-parallel ($\theta=\pi$). In these two cases, the string satisfies the Neuman-Neuman and Dirichlet-Dirichlet boundary conditions. For $\theta=\frac{\pi}{2}$, the string satisfies the Neuman-Dirichlet boundary condition and the vacuum energy increases to its maximum value $E=\frac{1}{24}\frac{\pi v}{l}$ which is a repulsive force [13]. Let us rewrite (57) as
\begin{equation}\label{1}
E=f(r)\frac{\pi v}{l}.
\end{equation}
As depicted in Fig. 2, the maximum of Casimir energy occurs for $r=\frac{1}{2}$.\\\\
\begin{figure}[H]
\centering
  \includegraphics[width=60mm]{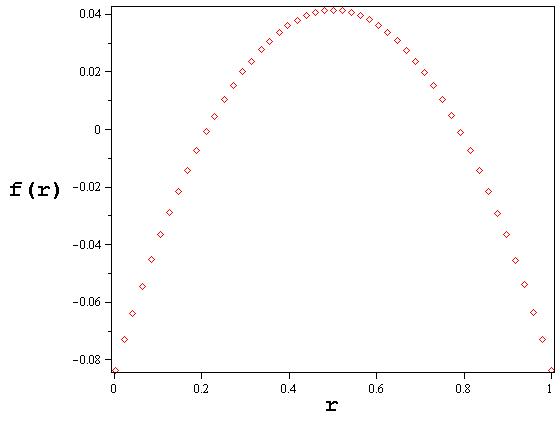}
  \caption{Variation of function $f(r)$ versus $r$. For $r=\frac{1}{2}$ Casimir energy reaches its maximum.}
\end{figure}
\appendix
\numberwithin{equation}{section}
\section{Appendix }
The infinite sum in formula (55) contains finite and divergent parts. To extract the finite term, we regularize it by introducing an exponential cut-off. Thus we write
{\setlength\arraycolsep{2pt}
\begin{eqnarray}
\nonumber
\lim_{\epsilon\rightarrow 0}\sum^{\infty}_{n=1}(n+r)e^{-\epsilon(n+r)}&=&\lim_{\epsilon\rightarrow 0}-\frac{d}{d\epsilon}\sum^{\infty}_{n=1}e^{-\epsilon(n+r)}\\\nonumber
&=&\lim_{\epsilon\rightarrow 0}-\frac{d}{d\epsilon}\,\frac{e^{-\epsilon(1+r)}}{1-e^{-\epsilon}}\\\nonumber
&=&\lim_{\epsilon\rightarrow 0}-\frac{d}{d\epsilon}\,\bigg[\frac{e^{-\epsilon(1+r)}}{\epsilon}\Big(1-\frac{1}{2}\epsilon+\frac{1}{6}\epsilon^2+\cdot\cdot\cdot\Big)^{-1}\bigg]\\\nonumber
&=&\lim_{\epsilon\rightarrow 0}-\frac{d}{d\epsilon}\,\bigg[e^{-\epsilon(1+r)}\Big(\frac{1}{\epsilon}+\frac{1}{2}+\frac{1}{12}\epsilon+\cdot\cdot\cdot\Big)\bigg]\\\nonumber
&=&\lim_{\epsilon\rightarrow 0}-\frac{d}{d\epsilon}\,\bigg\{\Big[1-\epsilon(1+r)+\frac{1}{2}\epsilon^2(1+r)^2+\cdot\cdot\cdot\Big]\Big(\frac{1}{\epsilon}+\frac{1}{2}+\frac{1}{12}\epsilon+\cdot\cdot\cdot\Big)\bigg\}\\\nonumber
&=&\lim_{\epsilon\rightarrow 0}-\frac{d}{d\epsilon}\,\bigg\{\Big[\frac{1}{12}-\frac{1}{2}(1+r)+\frac{1}{2}(1+r)^2\Big]\epsilon+\textrm{divergent part}\bigg\}\\
&=&\frac{1}{24}-\frac{1}{2}\bigg(r+\frac{1}{2}\bigg)^2+\textrm{divergent part}.
\end{eqnarray}}

\end{document}